\documentclass[9pt,shortpaper,twoside,web]{ieeecolor2}
\usepackage{generic}
\usepackage{amsmath,amssymb,amsfonts}
\usepackage{algorithmic}
\usepackage{algorithm} 
\usepackage{graphicx}
\usepackage{textcomp}
\usepackage{dsfont}
\usepackage{bbm}
\usepackage{float}
\usepackage{stfloats}
\usepackage{booktabs}   
\usepackage{colortbl}   
\usepackage{multirow}   
\usepackage{cite}

\usepackage{hyperref}
\hypersetup{
    colorlinks=true,
    linkcolor=blue,
    citecolor=blue,
    urlcolor=blue
}
\usepackage{xcolor}
\usepackage{cleveref}
\usepackage{tikz}
\usetikzlibrary{positioning}

\def\BibTeX{{\rm B\kern-.05em{\sc i\kern-.025em b}\kern-.08em
    T\kern-.1667em\lower.7ex\hbox{E}\kern-.125emX}}
\begin{document}
\title{SomaNet: Weakly Supervised Learning for Instance Soma Segmentation in 3D Electron Microscopy with Partial Annotations}

\author{Mohammad Khateri, Morteza Ghahremani, Jussi Tohka, Alejandra Sierra
\thanks{This work was supported in part by the Research Council of Finland (Grant \#361370); by the Flagship of Advanced Mathematics for Sensing, Imaging, and Modelling (Grant \#358944); by the Finnish Foundation for Technology Promotion; and by Finnish Cultural Foundation.}
\thanks{Mohammad Khateri, Jussi Tohka, and Alejandra Sierra are with the A. I. Virtanen Institute for Molecular Sciences, Faculty of Health Sciences, University of Eastern Finland, Finland (e-mail:  mohammad.khateri@uef.fi; jussi.tohka@uef.fi; alejandra.sierralopez@uef.fi).  Morteza Ghahremani is with the Munich Center for Machine Learning at the Technical University of Munich, Germany (e-mail: morteza.ghahremani@tum.de). } }

\maketitle

\begin{abstract}
Soma instance segmentation, i.e., identifying and delineating individual cell somas as distinct instances, is crucial for cellular analysis and connectomic reconstruction. Three-dimensional electron microscopy (3D EM) provides nanometer-scale resolution for capturing fine-grained soma morphology. However, dense instance-level manual annotation is prohibitively costly, limiting the scalability of fully supervised methods.
To address this challenge, we propose SomaNet, a weakly supervised framework for 3D EM soma instance segmentation under partial annotation constraints. SomaNet adopts a teacher--student learning paradigm tailored to partial labels. The teacher is trained using partially annotated data to generate pseudo-labels, while the student jointly learns from the partial ground-truth annotations and the generated pseudo-labels, progressively recovering dense instance segmentations.
To accurately delineate cell somas under limited supervision with varying instance counts, SomaNet incorporates affinity learning, which encourages high similarity within instances and low similarity across instance boundaries. Semantic-guided affinity decoding and 2D-to-3D reconstruction then produce volumetrically consistent 3D soma instances while preserving the large receptive fields of 2D backbones. The framework is architecture-flexible and supports diverse backbones, including vision transformers and foundation models, enabling direct transfer of pretrained visual representations to volumetric EM segmentation. Experiments on 3D EM brain datasets demonstrate that SomaNet achieves accurate and robust soma instance segmentation across regions with diverse soma morphologies under partial annotation. Code is available at \url{https://github.com/mkhateri/SomaNet}.

\end{abstract}


\section{Introduction}
\label{sec:introduction}
\IEEEPARstart{S}{oma}, the neuronal cell body housing the nucleus and major organelles, is central to neuronal communication and brain function~\cite{kandel2000principles}. Soma instance segmentation, i.e., identifying and delineating each soma as a distinct instance, is essential for analyzing neuronal morphology, spatial organization, and neural connectivity in neuroscience~\cite{kasthuri2015saturated}. While optical microscopy enables large-scale neuronal imaging, its diffraction-limited resolution cannot resolve the ultrastructural features required for connectomic reconstruction~\cite {lichtman2011big}.

Recent advances in high-throughput three-dimensional electron microscopy (3D EM) have enabled nanometer-scale imaging of brain tissue~\cite{hildebrand2017whole, khateri2024no}, revealing neuronal structures in unprecedented detail. Cell somas serve as key landmarks for neuronal circuit reconstruction~\cite{motta2019dense}, and are also associated with cell-type classification and morphological quantification~\cite{jiang2015principles, lee2016anatomy}. However, the massive data volumes and structural intricacies of 3D EM make manual soma instance segmentation prohibitively costly at scale.

Deep learning has achieved remarkable success in scalable and accurate instance segmentation across various domains \cite{xu2017gland, minaee2021image, rettenberger2023self, lalit2022embedseg, archit2025segment, xiao2025protomgnet}. However, for soma segmentation in 3D EM, acquiring densely annotated training data is highly time-intensive and requires considerable domain expertise, limiting the feasibility of fully supervised approaches. Recent advances in computer vision offer promising alternatives by enabling automated instance segmentation with minimal supervision and reduced annotation requirements~\cite{han2024deep}.

Semi-supervised learning (SSL) addresses limited annotation by combining a small set of densely labeled examples with a larger pool of unlabeled data~\cite{zhang2023multi, peng2020deep, zhao2023augmentation}. However, in 3D EM applications even a few such dense annotations are rarely available; labels are typically partial, covering only a subset of instances and posing a distinct supervision challenge. Weakly supervised learning (WSL) is better suited to this regime, operating directly on coarse, sparse, or partial annotations~\cite{chen2024spatial, nishimura2021weakly, zhang2023weakly, lerousseau2020weakly} without requiring any densely labeled examples, offering a more scalable and annotation-efficient solution.

Despite growing success in segmentation tasks~\cite{zhang2021weakly, chinkamol2022octave}, WSL remains largely underexplored for instance segmentation under partial annotations, particularly for soma instance segmentation in 3D EM, where dense cellular packing, irregular morphologies, and incomplete labels pose considerable challenges. To date, only one notable study has addressed soma instance segmentation in 3D EM \cite{liu2023soma}.  Although it operates on partially labeled data, it was not explicitly designed for weak supervision. The method employs a conventional proposal-based architecture, i.e., a detection-then-segmentation pipeline, and does not directly address the limitations introduced by partial instance-level annotations. Such approaches often struggle with irregularly shaped somas, particularly in high-resolution EM volumes where object extents may exceed the model’s receptive field, highlighting the need for methods that are both robust to complex morphology and capable of learning from partial supervision.

In parallel, foundation vision models such as DINOv2~\cite{oquab2023dinov2} have demonstrated powerful feature representations that generalize across diverse visual tasks. Despite this potential, their application to weakly supervised instance segmentation of soma in 3D EM remains unexplored, presenting an untapped opportunity to advance segmentation under limited supervision.

To this end, we propose SomaNet, a proposal-free, weakly supervised framework for soma instance segmentation in 3D EM that infers dense instance segmentations from partial annotations. A teacher--student paradigm leverages consistency regularization and exponential moving average stabilization to propagate partial supervision and progressively refine pseudo-labels, while affinity learning embeds pixels into a similarity space that promotes intra-instance coherence and sharp inter-instance boundaries. To reconcile large in-plane receptive fields with volumetric consistency, the resulting high-quality 2D instance predictions are assembled into volumetrically consistent 3D soma instances through geometric cross-slice association and volumetric refinement. The framework is architecture-flexible, supporting a wide range of backbone architectures, including vision transformers and foundation models, thereby enabling seamless transfer of pretrained visual representations to the EM domain. Our main contributions are as follows:

\begin{itemize}
\item We present SomaNet, a proposal-free, weakly supervised framework for soma instance segmentation in 3D EM that infers dense instance labels from partial annotations.

\item We design a teacher–student paradigm with affinity-based embeddings to propagate partial supervision into dense, boundary-accurate segmentations at a fraction of the annotation cost.

\item We provide theoretical analysis showing that mask loss induces biased gradients under partial annotation, whereas affinity learning yields gradients that are less sensitive to this bias.

\item We demonstrate that SomaNet is flexible across architectures, including Vision Transformers and foundation models.
\end{itemize}

\noindent The remainder of this manuscript is organized as follows: Section II reviews related work; Section III details the proposed method; Section IV describes the experimental setup and presents the results; and Section V concludes the paper.

\section{Related Works}
\subsection{Instance Segmentation}
Instance segmentation approaches are broadly classified as proposal-based or proposal-free. Proposal-based methods first detect object regions, typically in the form of bounding boxes, followed by segmentation~\cite{liu2021panoptic, he2017mask, liu2019nuclei}. While effective in structured settings, they often struggle in cluttered scenes with overlapping or irregularly shaped objects, particularly in high-resolution biomedical imagery where object extents may exceed the model’s receptive field~\cite{liu2025graph}.

Proposal-free methods bypass bounding-box proposals by learning instance-aware representations directly~\cite{lalit2022embedseg, huang2022learning, de2017semantic}. They typically predict intermediate representations, such as boundaries~\cite{kirillov2017instancecut}, affinities~\cite{liu2018affinity}, embeddings~\cite{de2017semantic}, or semantic maps~\cite{abdollahzadeh2021deepacson}, which are then clustered or decoded into individual instances during post-processing.

Metric and affinity learning are two key proposal-free strategies. Metric learning constructs an embedding space where features from the same instance are pulled together while those from different instances are pushed apart~\cite{de2017semantic}. Although effective for promoting inter-instance separability, it often lacks spatial coherence, particularly in crowded or morphologically ambiguous regions. In contrast, affinity learning predicts pairwise relationships between neighboring pixels or voxels~\cite{gao2019ssap, wang2023object, briggman2009maximin}, enabling precise boundary delineation and strong intra-instance consistency~\cite{lee2017superhuman}. By modeling spatial continuity rather than explicit instance identity, affinity learning is easier to train, less sensitive to annotation sparsity and boundary ambiguity, and remains effective even with limited labels~\cite{tu2018learning, wang2023object}, making it well suited for complex and irregular structures such as somas in 3D EM.

\subsection{Segmentation under Limited Supervision}
To reduce reliance on dense annotations, recent work has turned to learning with limited supervision. SSL addresses this by combining a small labeled subset with a larger pool of unlabeled data~\cite{liu2025certainty}, using techniques such as consistency regularization~\cite{wu2024instance, ke2020guided}, pseudo-labeling~\cite{berthelot2019mixmatch}, data augmentation~\cite{berthelot2019remixmatch}, and label propagation~\cite{bengio2006label}. WSL, in contrast, operates under settings where supervision is provided in the form of coarse, sparse, or partial annotations rather than fully labeled data~\cite{zhang2020weakly, gao2022segmentation, zhou2018weakly, lerousseau2020weakly}. This makes WSL particularly suitable for 3D EM, where dense annotation is often infeasible.

Among these strategies, teacher--student frameworks~\cite{tarvainen2017mean, tan2026towards, filipiak2024polite} have emerged as a powerful paradigm, enabling a student model to learn from pseudo-labels generated by a teacher and effectively propagate information from labeled to unlabeled regions. While successful in classification, semantic segmentation, and more recently instance segmentation, their application to weakly supervised 3D EM soma instance segmentation remains largely unexplored. In addition, effectively exploiting vision foundation models within such frameworks represents a promising direction for weakly supervised 3D EM soma instance segmentation.

\subsection{Soma Instance Segmentation in 3D EM}

\subsubsection{Neuron- and Nucleus-based Approaches}
Early EM segmentation methods focused on reconstructing entire neurons using architectures such as flood-filling networks~\cite{januszewski2018high} and MALA~\cite{funke2018large}. While effective for tracing neuronal arbors, these approaches do not explicitly model somas, often failing to separate them from neurites and yielding coarse delineations. Their high computational cost and complex post-processing limit scalability for large-scale analysis.

Given the anatomical relationship between somas and nuclei, nucleus segmentation has also been explored as a proxy for soma detection~\cite{mu20213d, lin2021nucmm}. However, because nuclei occupy only a subset of the soma volume, they cannot capture the full extent and morphological variability of neuronal somata, particularly those with irregular morphologies.

\vspace{-0.2mm}

\subsubsection{Dedicated Soma Segmentation}
To address these limitations, Liu et al.~\cite{liu2023soma} introduced the first dedicated soma instance segmentation pipeline, built on the EMADS dataset of adult \textit{Drosophila} brain. Their two-stage method combines seed-based detection with patch-wise refinement. However, the approach relies on a proposal-based architecture that depends on accurate first-stage detection, making it less robust for irregularly shaped somas. Moreover, it is not designed for weak supervision, which limits its applicability in partially annotated settings.

\vspace{-0.3mm}
\section{Proposed Method}
\label{sec:method}

Fully supervised soma instance segmentation in 3D EM requires dense annotations that are impractical at scale. To address this challenge, we propose SomaNet, a weakly supervised framework that infers dense soma instance segmentations from partial annotations through teacher--student learning with affinity learning, enabling accurate and scalable 3D soma instance segmentation from partial annotations. The following subsections describe each component of the proposed framework.

\vspace{-0.3mm}

\subsection{Affinity Learning}
\label{subsec:affinity}

Affinity learning models pairwise relationships between neighboring pixels and captures instance boundaries~\cite{funke2018large,huang2022learning}. Unlike direct affinity prediction, which predicts affinity maps using CNNs~\cite{maire2016affinity,gao2019ssap}, we adopt an explicit pixel embedding approach that derives affinities from learned embeddings~\cite{huang2022learning}. This formulation preserves instance-level semantic information. Given an image $I\in\mathbb{R}^{C\times H\times W}$\footnote{$C=1$ for single-channel EM data.}, the network maps it into an embedding space $E\in\mathbb{R}^{D\times H\times W}$, where $D$ denotes the embedding dimension. For affinity computation, embeddings are $\ell_2$-normalized along the channel dimension. To capture both fine- and coarse-scale boundaries, we define a set of affinity ranges $R$ (e.g., $R=\{1,3,5,9,11,19,27,35\}$). For each range, affinities are computed between each pixel and its $N$ neighboring locations, where $N$ denotes the number of affinity directions (here $N=4$, the forward offsets of an 8-connected neighborhood). Let $\mathcal N(v)$ denote the set of neighboring pixels associated with the selected affinity ranges $R$ and affinity directions $N$. The predicted affinity tensor $\hat A\in\mathbb{R}^{C_{\mathrm{aff}}\times H\times W}$, where $C_{\mathrm{aff}}=|R|\times N$, is computed as
\begin{equation}
\hat A_{v,k}=e_v^\top e_k,
\qquad
k\in\mathcal N(v),
\end{equation}
where $e_v\in\mathbb{R}^D$ denotes the embedding at pixel $v$, and $k$ is one of the neighboring pixels in $\mathcal N(v)$. The corresponding ground-truth affinity tensor $A$, derived from the instance label map $y\in\mathbb{Z}^{H\times W}$, is defined as
\begin{equation}
\label{eq:affinity}
A_{v,k}=
\begin{cases}
1,& y_v=y_k,\\
0,& \text{otherwise},
\end{cases}
\qquad
k\in\mathcal N(v),
\end{equation}
where an affinity value of $1$ indicates that pixels $v$ and $k$ belong to the same instance, whereas a value of $0$ indicates that they belong to different instances.

\begin{figure*}[!ht]
    \centering
    \includegraphics[width=0.87\linewidth]{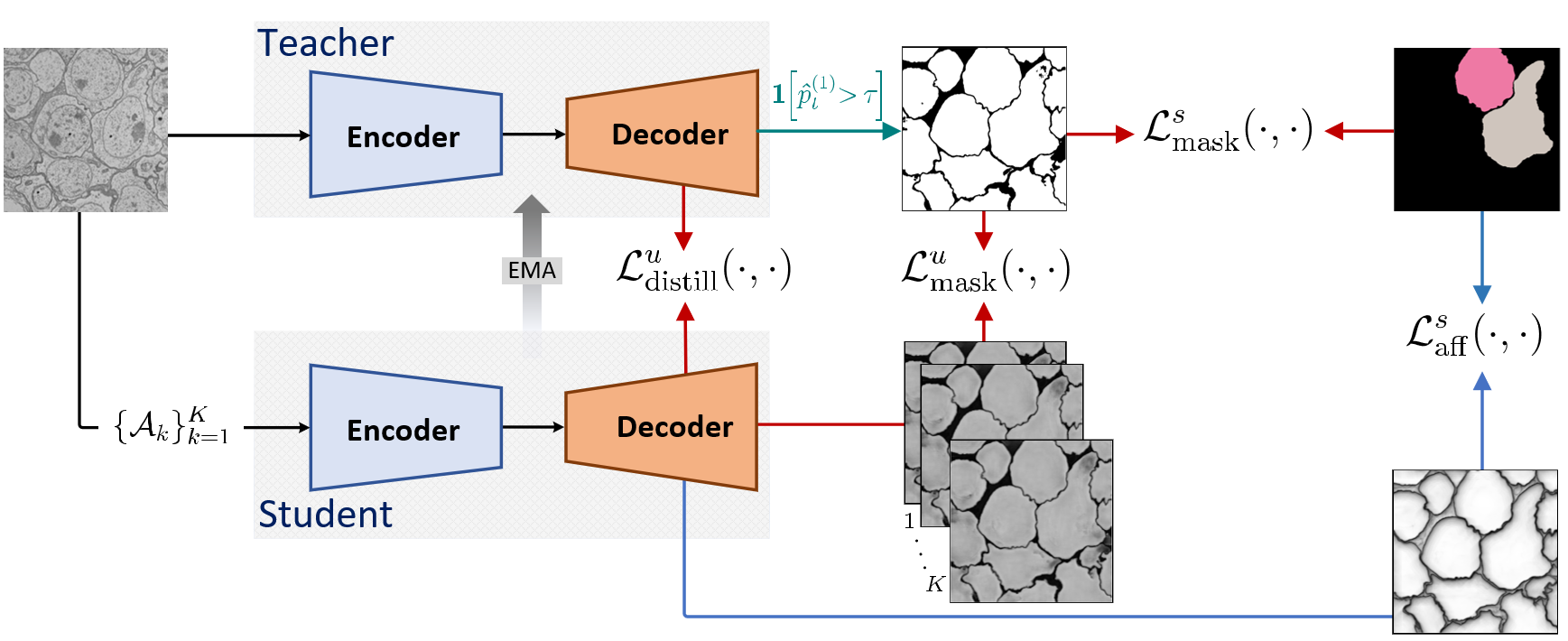}
    \caption{
    Overview of the proposed teacher--student framework. The teacher (top) processes the original image to generate pseudo-labels $\tilde{y}_j = \mathbbm{1}[\hat{p}_t^{(1)} > \tau]$. The student (bottom) receives $K$ augmented views and is trained with supervised and consistency objectives. Supervision on labeled regions is applied via $\mathcal{L}_{\mathrm{sup}}$, combining mask and affinity losses. Consistency is enforced through pseudo-labels via $\mathcal{L}_{\mathrm{mask}}^{u}$, while feature-level alignment between teacher and student is imposed by $\mathcal{L}_{\mathrm{distill}}$. The teacher parameters are updated as an exponential moving average of the student parameters. The affinity visualization shows the mean over the nearest-neighbor offset channels in $R$, where brighter regions indicate higher predicted similarity between adjacent pixels and darker regions correspond to instance boundaries.
    }
    \label{fig:teacher-student}
\end{figure*}
\subsection{Supervised Loss Formulation}
\label{subsec:wsl}

Let $\mathcal{D}_\ell = \left\{ (x_i, y_i, \Omega_i) \right\}_{i=1}^{n}$ denote the labeled dataset, where $x_i \in \mathbb{R}^{H \times W}$ is an input image, $y_i \in \mathbb{Z}^{H \times W}$ is the ground-truth instance map, and $\Omega_i \subseteq \mathcal{V} = \{1, \dots, H \} \times \{1, \dots, W \}$ denotes the set of annotated pixels, with $\Omega_i^u = \mathcal{V} \setminus \Omega_i$ the unannotated pixels.
Directly supervising instance identities is impractical when thousands of instances are present and annotations are partial. Instead, we decompose the task into two complementary objectives: (i) \textbf{Mask loss}, which enforces semantic foreground--background separation, and (ii) \textbf{Affinity loss}, which promotes instance-level separation via embedding similarity. The model $f_\theta$ outputs a segmentation mask $\hat{y} \in [0,1]^{H \times W}$ and pixel-wise embeddings $e \in \mathbb{R}^{D \times H \times W}$.
Under partial annotation, the mask loss is computed over $\mathcal{V}$ with $y(v)$ as a binary foreground indicator, set to zero for $v \in \Omega_i^u$:
\begin{equation}
\label{eq:mask_loss}
\mathcal{L}_{\mathrm{mask}} = \frac{1}{|\mathcal{V}|} \sum_{v \in \mathcal{V}} \mathrm{CE}\left(y(v), \hat{y}(v)\right),
\end{equation}
where $\mathrm{CE}(\cdot, \cdot)$ denotes the cross-entropy loss.

The affinity loss supervises the embedding-based affinities using a mean squared error loss~\cite{huang2022learning}:
\begin{equation}
\label{eq:aff_loss}
\mathcal{L}_{\mathrm{aff}} = \frac{1}{|\mathcal{V}|} \sum_{v \in \mathcal{V}} \sum_{k \in \mathcal{N}(v)} \left( A_{v,k} - e_v^\top e_k \right)^2,
\end{equation}
where $A_{v,k}$ is the ground-truth affinity defined in Eq.~(\ref{eq:affinity}).
The total supervised loss is:
\begin{equation}
\label{eq:sup_loss}
\mathcal{L}_{\mathrm{sup}} = \lambda_{\mathrm{mask}} \mathcal{L}_{\mathrm{mask}} + \lambda_{\mathrm{aff}} \mathcal{L}_{\mathrm{aff}}.
\end{equation}

The $\mathcal{L}_{\mathrm{mask}}$ provides the foreground signal needed for semantic separation but incurs a false-negative bias under partial annotations, while $\mathcal{L}_{\mathrm{aff}}$ provides boundary supervision via local pairwise relationships and is less biased to partial annotation than $\mathcal{L}_{\mathrm{mask}}$ (Appendix~\ref{app:affinity_robustness}).
\subsection{Teacher–Student Training}
\label{subsec:teacher_student}

To exploit unlabeled regions arising from partial annotations, we employ a teacher--student framework in which the teacher generates pseudo-labels from the input, and the student is trained to match them under stochastic augmentations of the same input, in addition to ground-truth supervision, enforcing pixel-wise consistency. The teacher parameters are updated as an exponential moving average (EMA) of the student, providing stable and progressively refined targets during training (Figure~\ref{fig:teacher-student}).

\subsubsection{Architecture and Initialization}
Both teacher and student networks share an identical architecture and are initialized from a model pretrained on labeled regions using $\mathcal{L}_{\mathrm{sup}}$ (Eq.~\ref{eq:sup_loss}). The teacher $f_{\theta_t}$ is updated as an EMA of the student parameters:
\begin{equation}
\label{eq:ema}
    \theta_t \leftarrow \alpha \, \theta_t + (1 - \alpha) \, \theta_s,
\end{equation}
where $\alpha \in [0,1)$ controls the momentum, ensuring the teacher provides stable, temporally smoothed predictions during training.

\subsubsection{Pseudo-Label Generation}
The teacher generates foreground pseudo-labels by thresholding its softmax outputs. Let $\hat{p}_t = \mathrm{softmax}(f_{\theta_t}(x_j))$ denote the teacher probability map. A binary pseudo-label is derived as:
\begin{equation}
\label{eq:pseudo}
    \tilde{y}_j = \mathbbm{1}\left[\hat{p}_t^{(1)} > \tau\right],
\end{equation}
where $\hat{p}_t^{(1)}$ is the foreground probability, $\tau$ is a confidence threshold, and $\mathbbm{1}[\cdot]$ denotes the indicator function.

\subsubsection{Student Training}
The student $f_{\theta_s}$ is trained using both ground-truth supervision and consistency regularization. For labeled data, supervision follows Eq.~\ref{eq:sup_loss}. In addition, the student is trained to match the teacher's predictions under $K$ stochastic augmentations $\{\mathcal{A}_k\}_{k=1}^{K}$ that preserve spatial correspondence:
\begin{equation}
\label{eq:mask_unsup}
    \mathcal{L}_{\mathrm{mask}}^{u} = \frac{1}{K|\mathcal{V}|} \sum_{k=1}^{K} \sum_{v \in \mathcal{V}} \text{CE}\left(\tilde{y}_j(v),\, f_{\theta_s}(\mathcal{A}_k(x_j))(v)\right).
\end{equation}
We further align student and teacher embeddings via a feature-level distillation loss:

\begin{equation} 
\label{eq:distill}
\mathcal{L}_{\mathrm{distill}} = \frac{1}{|\mathcal{V}|} \sum_{v \in \mathcal{V}} \| e_s(v) - e_t(v) \|_2^2,   
\end{equation} 
where $e_s(\cdot)$ and $e_t(\cdot)$ are the student and teacher embeddings. 

\subsubsection{Overall Objective}
The total consistency loss combines mask and feature terms:
\begin{equation}
\label{eq:cons}
    \mathcal{L}_{\mathrm{cons}} = \lambda_{\mathrm{mask}} \mathcal{L}_{\mathrm{mask}}^{u} + \lambda_{\mathrm{distill}} \mathcal{L}_{\mathrm{distill}}.
\end{equation}

\begin{figure}[!t]
    \centering
    \includegraphics[width=0.99\linewidth]{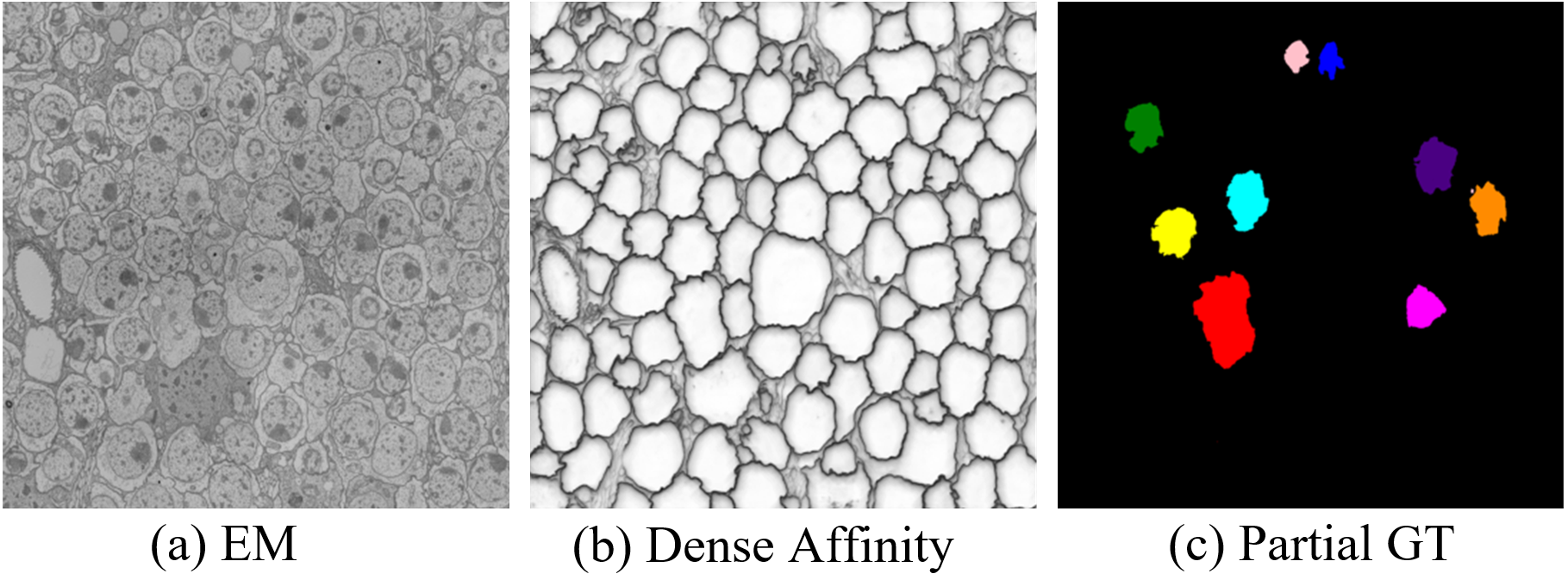}
    \caption{
    Propagation of partial annotations into dense affinity during SomaNet training. (a) Raw EM input, (b) dense affinity predicted by SomaNet, and (c) partially annotated ground truth covering only a subset of soma instances. The proposed framework enables dense prediction from limited annotations.
    }
    \label{fig:partial2dense}
\end{figure}

The overall objective integrates supervised and consistency losses:
\begin{equation}
\label{eq:total}
    \mathcal{L}_{\mathrm{total}} = \mathcal{L}_{\mathrm{sup}} + \lambda_u \mathcal{L}_{\mathrm{cons}},
\end{equation}
where $\lambda_u$ controls the contribution of the consistency term, enabling progressive refinement: the teacher provides increasingly reliable pseudo-labels through EMA updates, improving generalization under partial annotation settings (Figure~\ref{fig:partial2dense}).

\subsection{2D-to-3D Instance Segmentation}
\label{subsec:2d_to_3d}

Given the trained teacher-student model, SomaNet converts the predicted semantic and affinity maps into volumetrically consistent 3D soma instances. For each EM slice, the semantic prediction localizes foreground regions, while the affinity maps encode pairwise pixel connectivity to separate adjacent soma cells. These complementary cues are integrated through semantic-guided affinity decoding, where the semantic prediction suppresses non-soma regions and constrains the affinity-guided seeded watershed to the foreground, while the affinity maps separate adjacent somas. A subsequent boundary-aware agglomeration further refines the resulting 2D soma instances.

The resulting slice-wise instances are assembled into 3D objects through geometric cross-slice association based on inter-slice mask overlap. Consecutive slices are linked using one-to-one instance correspondences, while disconnected fragments are re-linked across short gaps to recover temporarily missing detections. Finally, volumetric refinement restores short missing segments, enforces slice-to-slice consistency, removes small spurious fragments, and relabels connected components to produce the final 3D instance segmentation. The complete SomaNet pipeline is summarized in Algorithm~\ref{alg:somanet}.

\subsection{Adapting Backbone Architectures in SomaNet}

\begin{figure}[!t]
    \centering
    \includegraphics[width=0.85\linewidth,height=0.35\linewidth]{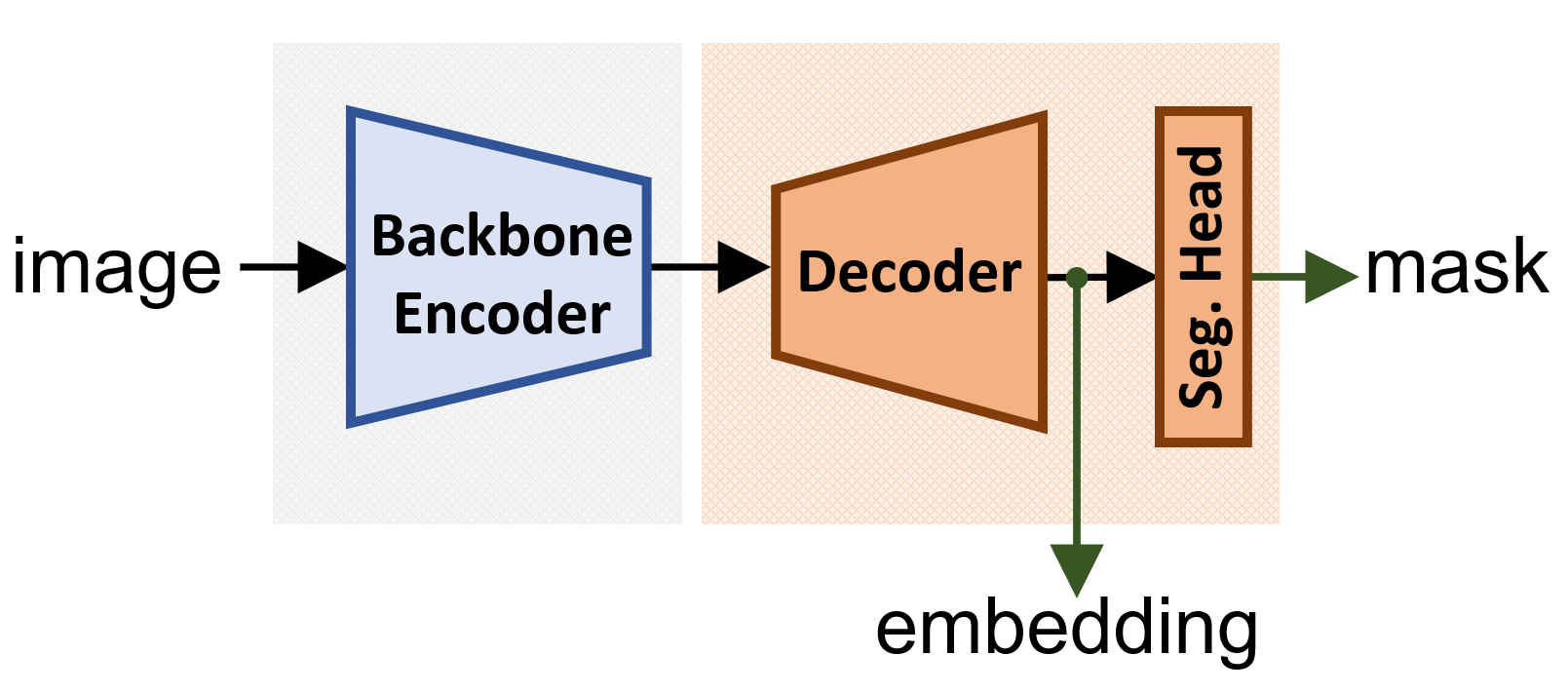}
    \caption{Backbone adaptation in SomaNet. Existing encoder--decoder architectures are adapted with minimal modifications. The decoder produces a shared embedding representation that is used for affinity learning to learn discriminative voxel features for instance segmentation, while the same decoder features are simultaneously provided to a semantic segmentation head for foreground mask prediction.}
    \label{fig:architecture}
\end{figure}

SomaNet is designed to be architecture-flexible and can be integrated with diverse 2D encoder--decoder backbones using only lightweight modifications. Specifically, the original task-specific prediction head is replaced with a semantic segmentation head. The rich decoder features are directly reused as a shared embedding for affinity learning, enabling discriminative learning of voxel representations for instance segmentation, while the same decoder features are simultaneously used by the semantic segmentation head for foreground mask prediction (Figure~\ref{fig:architecture}). To demonstrate this flexibility, we adapt two representative backbone architectures.

First, we employ DINOv2 (ViT-S/14) \cite{oquab2023dinov2} as the encoder to leverage its powerful pretrained visual representations. Inspired by the design principles of ViT-Adapter~\cite{chen2022vision}, we augment DINOv2 with a lightweight dense prediction decoder comprising a convolutional spatial-prior branch, bidirectional cross-attention modules for feature interaction across multiple transformer depths, a full-resolution convolutional branch that preserves fine spatial details lost during the $14\times14$ patch embedding, and residual convolutional refinement blocks. The refined decoder features are then used as the shared embedding for affinity learning and simultaneously provided to the semantic segmentation head for foreground mask prediction. The DINOv2 backbone is initialized with publicly available pretrained weights. During training, only the final four transformer blocks are fine-tuned, while the remaining backbone layers remain frozen. 

Second, we adapt the SwinIR \cite{liang2021swinir} architecture, which naturally preserves dense spatial correspondence through shallow convolutional feature extraction followed by residual Swin Transformer blocks. As SwinIR was originally designed for image restoration, we remove its image reconstruction head and long skip connection. The resulting decoder features are then used as the shared embedding for affinity learning and simultaneously provided to the semantic segmentation head for foreground mask prediction.

\begin{algorithm}[t]
\caption{Training and inference procedure of SomaNet}
\label{alg:somanet}

\begin{algorithmic}[1]

\REQUIRE Partially annotated dataset $\mathcal{D}_{\ell}$ and initial weights $\theta_0$ (pretrained or random)
\ENSURE Final 3D instance segmentation $Y_{\mathrm{Pred}}$

\vspace{3pt}
\STATE \textcolor{gray}{\textit{\# Stage 1: Supervised pretraining/fine-tuning on partial labels}}

\STATE $\theta \gets \theta_0$

\FOR{$t = 1,\ldots,T_1$}
    \STATE $(\hat{y}_i,e_i)\gets f_{\theta}(x_i)$
    \STATE Update $\theta$ by optimizing $\mathcal{L}_{\mathrm{sup}}$
    \hfill{\scriptsize Eq.~(\ref{eq:sup_loss})}
\ENDFOR

\STATE $\theta_s\gets\theta$;\quad $\theta_t\gets\theta$

\vspace{3pt}
\STATE \textcolor{gray}{\textit{\# Stage 2: Teacher--student consistency training}}

\FOR{$t = 1,\ldots,T_2$}

    \STATE \textbf{Teacher forward (no gradient):}
    \STATE \quad $\hat{p}_t\gets\mathrm{softmax}(f_{\theta_t}(x_i))$
    \STATE \quad $\tilde{y}_i\gets\mathbbm{1}\!\left[\hat{p}_t^{(1)}>\tau\right]$

    \STATE \textbf{Student forward on $K$ augmented views:}
    \STATE \quad $(\hat{y}_i^{(k)},e_i^{s,(k)})\gets f_{\theta_s}(\mathcal{A}_k(x_i))$,\quad $k=1,\ldots,K$

    \STATE $\mathcal{L}\gets\mathcal{L}_{\mathrm{sup}}
    +\lambda_u\mathcal{L}_{\mathrm{cons}}$
    \hfill{\scriptsize Eq.~(\ref{eq:total})}

    \STATE Update $\theta_s$ by optimizing $\mathcal{L}$

    \STATE $\theta_t\gets\alpha\theta_t+(1-\alpha)\theta_s$
    \hfill{\scriptsize Eq.~(\ref{eq:ema})}

\ENDFOR

\vspace{3pt}
\STATE \textcolor{gray}{\textit{\# Stage 3: Inference and 3D instance reconstruction}}

\STATE Disable gradient computation

\FOR{each EM slice $x_z$ in the input volume}

    \STATE $(\hat{y}_z,e_z)\gets f_{\theta_t}(x_z)$

    \STATE $Y_{\mathrm{Pred}}^{(z)}
    \gets
    \mathrm{InstanceDecode}(\hat{y}_z,e_z)$

\ENDFOR

\STATE $Y_{\mathrm{Pred}}
\gets
\mathrm{2Dto3DInstanceSeg}
\!\left(
\{Y_{\mathrm{Pred}}^{(z)}\}
\right)$

\RETURN $Y_{\mathrm{Pred}}$

\end{algorithmic}
\end{algorithm}

\section{Experiments}

\subsection{Dataset}
We evaluate our method on the EMADS dataset~\cite{liu2023soma}, the only publicly available 3D instance-labeled dataset of neuronal somas from an adult Drosophila brain. 
EMADS is derived from the full adult fly brain dataset~\cite{zheng2018complete}, acquired via serial-section transmission electron microscopy (ssTEM).
The data were downsampled to a resolution of $16 \times 16 \times 40$\,$nm^3$ and partitioned into 3D blocks of size $1836 \times 1836 \times 186$ voxels. Ten spatially distributed blocks, sampled from regions with varying soma densities, were partially annotated at the voxel level, yielding 204 soma instances (approximately 20 per block) with diverse morphologies and over $8 \times 10^9$ labeled voxels. In addition, two anatomically distinct regions were almost densely annotated, producing sub-volumes of $650 \times 650 \times 550$ voxels each. These test volumes encompass diverse tissue characteristics, soma morphologies, and imaging artifacts, providing a benchmark for evaluating instance segmentation frameworks.

\subsection{Evaluation Metrics}
We evaluate soma instance segmentation performance using region-overlap metrics (Dice and IoU) and instance-aware metrics (PQ, ARAND, and VOI).
\subsubsection{Dice Score}
The Dice coefficient~\cite{taha2015metrics} measures the volumetric overlap between the predicted and ground-truth masks:
\begin{equation}
    \mathrm{Dice} =
    \frac{2|Y_{\mathrm{Pred}} \cap Y_{\mathrm{GT}}|}
    {|Y_{\mathrm{Pred}}| + |Y_{\mathrm{GT}}|}.
\end{equation}
Higher values indicate better segmentation quality.
\subsubsection{Jaccard Index (IoU)}
The Jaccard index~\cite{taha2015metrics}, also known as the Intersection-over-Union (IoU), measures the volumetric overlap between the predicted and ground-truth masks:
\begin{equation}
    \mathrm{IoU} =
    \frac{|Y_{\mathrm{Pred}} \cap Y_{\mathrm{GT}}|}
    {|Y_{\mathrm{Pred}} \cup Y_{\mathrm{GT}}|}.
\end{equation}
Higher values indicate better segmentation quality.

\subsubsection{Panoptic Quality (PQ)}
The PQ~\cite{kirillov2019panoptic} jointly evaluates instance recognition and segmentation quality by matching predicted and ground-truth instances using one-to-one matching with an IoU threshold of 0.5. It is defined as
\begin{equation}
\mathrm{PQ}=\frac{\sum_{(p,g)\in TP}\mathrm{IoU}(p,g)}{|TP|+\frac{1}{2}|FP|+\frac{1}{2}|FN|},
\end{equation}
where $TP$ denotes the set of matched instance pairs, while $FP$ and $FN$ denote the unmatched predicted and ground-truth instances. Higher values indicate better instance segmentation quality.
\subsubsection{Adapted Rand Error (ARAND)}
The ARAND~\cite{arganda2015crowdsourcing} measures the disagreement between the predicted and ground-truth instance segmentations based on pairwise voxel assignments. It is defined as the complement of the pairwise Rand index F-score:
\begin{equation}
\mathrm{ARAND} = 1- \frac{2\sum_{i,j} n_{ij}^{2}} {\sum_i a_i^{2}+\sum_j b_j^{2}},
\end{equation}
where $n_{ij}$ denotes the number of voxels shared by predicted instance $i$ and ground-truth instance $j$, and $a_i=\sum_j n_{ij}$ and $b_j=\sum_i n_{ij}$ denote the corresponding marginal voxel counts. Lower values indicate better instance segmentation quality.

\subsubsection{Variation of Information (VOI)}
The VOI~\cite{taha2015metrics} measures the information-theoretic distance between the predicted and ground-truth instance segmentations:
\begin{equation}
\mathrm{VOI}=\underbrace{H(Y_{\mathrm{Pred}} \mid Y_{\mathrm{GT}})}_{\mathrm{VOI}_{\mathrm{split}}}
+\underbrace{H(Y_{\mathrm{GT}} \mid Y_{\mathrm{Pred}})}_{\mathrm{VOI}_{\mathrm{merge}}}.
\end{equation}
where $H(\cdot\mid\cdot)$ denotes the conditional entropy, $\mathrm{VOI}_{\mathrm{split}}$ measures over-segmentation (split) errors, and $\mathrm{VOI}_{\mathrm{merge}}$ measures under-segmentation (merge) errors. Lower values indicate better instance segmentation quality.

\subsection{Implementation Details}
\paragraph{Training}
Our method is implemented in PyTorch with multi-GPU parallel processing and ran on two NVIDIA A100 GPUs. Training uses bfloat16 mixed-precision to reduce memory consumption while maintaining numerical stability, with BatchNorm layers kept in float32; for float16 training, gradient scaling is applied via PyTorch's GradScaler. We extract image patches of size $400 \times 400$ from full-resolution EM volumes, and augment them in two tiers: rotation-sensitive geometry (flips, rotation, rescaling) is applied once per patch to keep the teacher and student spatially aligned, then rotation-insensitive appearance transforms, e.g., Gaussian noise, blur, grayscale, cut-out, and synthetic line artifacts, are applied stochastically to student inputs. Teacher-student models are trained using the Adam optimizer with a base learning rate of $10^{-4}$ and a batch size of $2$ for approximately two days. The supervised loss combines embedding-driven affinity and mask terms with weights $\lambda_{\mathrm{aff}} = 0.005$ and $\lambda_{\mathrm{mask}} = 1.0$, where the mask loss is cross-entropy. For weakly-supervised learning, feature distillation uses an MSE loss with weight $\lambda_{\mathrm{distill}} = 0.1$, and the overall weakly-supervised contribution is scaled by $\lambda_u = 0.5$. The EMA coefficient for updating the teacher network is $\alpha = 0.999$, with both teacher and student initialized from models pretrained on labeled regions; the teacher remains frozen during training and is updated only via EMA, using a pseudo-label threshold of $\tau = 0.5$. Affinity maps use multi-scale spatial offsets $R \in \{1, 3, 5, 9, 11, 19, 27, 35\}$, with $N=4$, and class imbalance is addressed via weighted cross-entropy with weights inversely proportional to class frequency.

\paragraph{Inference}
At inference, each 2D EM slice is processed independently to predict a semantic foreground probability map and multi-scale affinity embeddings. The size-agnostic SwinIR processes full slices, whereas DINOv2 is applied to overlapping $400\times400$ tiles (stride 150) fused using Gaussian weighting ($\sigma=100$ pixels). The semantic map is binarized using $\tau=0.3$, followed by a morphological opening ($5\times5$) and removal of foreground components smaller than $250$ pixels. Multi-scale affinity maps over $R$, gated by the semantic prediction, drive an affinity-guided seeded watershed followed by boundary-aware agglomeration (boundary threshold 0.35) to generate 2D instances. Adjacent slices are linked using one-to-one geometric matching with a minimum overlap of $20$ pixels and re-linking across gaps of up to five slices. Finally, volumetric refinement fills brief interior holes and discards instances spanning fewer than three slices.

\vspace{-2mm}
\subsection{Comparison}
We compare SomaNet against strong baselines derived from the EMADS framework~\cite{liu2023soma}, currently the only dedicated method for soma instance segmentation in volumetric EM. EMADS is instantiated with two representative backbone architectures, 3D U-Net and Swin UNETR~\cite{tang2022self}, using the authors' publicly available code and pretrained weights. In addition, we evaluate SomaNet across two backbone architectures to demonstrate its generality: SwinIR~\cite{liang2021swinir} (transformer-based) and DINOv2~\cite{oquab2023dinov2} with an adapter (vision foundation model). This setup enables comparison across both method design and architectural capacity. Evaluation is performed on the two designated test blocks of the EMADS benchmark~\cite{liu2023soma}. Although these blocks are nearly fully annotated, somas intersecting volume boundaries are not included in the annotated ground truth, as they do not form complete 3D structures. All methods are therefore evaluated only within annotated ground-truth regions.

\begin{table*}[t]
\centering
\caption{Quantitative comparison of soma instance segmentation methods on 3D EM test sets. Best results are highlighted in \textbf{bold}.}
\label{tab:method_comparison}

\resizebox{0.82\textwidth}{!}{%
\begin{tabular}{llccccccc}
\toprule
\rowcolor{gray!15}
\textbf{Dataset} & \textbf{Method}
& \textbf{Dice$\uparrow$}
& \textbf{IoU$\uparrow$}
& \textbf{PQ$\uparrow$}
& \textbf{ARAND$\downarrow$}
& \textbf{VOI$_{\mathrm{total}}\downarrow$}
& \textbf{VOI$_{\mathrm{split}}\downarrow$}
& \textbf{VOI$_{\mathrm{merge}}\downarrow$} \\
\midrule

\multirow{4}{*}{\textbf{Test Block 1}}
& EMADS (3D U-Net)
& 0.835 & 0.717 & 0.643 & 0.737 & 2.417 & 0.853 & 1.564 \\

& EMADS (Swin UNETR)
& 0.742 & 0.590 & 0.625 & 0.822 & 2.723 & 0.657 & 2.066 \\

& SomaNet (DINOv2) [Ours]
& \textbf{0.993} & \textbf{0.986} & 0.706 & 0.104 & 0.411 & 0.244 & \textbf{0.166} \\

& SomaNet (SwinIR) [Ours]
& 0.991 & 0.982 & \textbf{0.767} & \textbf{0.103} & \textbf{0.454} & \textbf{0.220} & 0.233 \\

\midrule

\multirow{4}{*}{\textbf{Test Block 2}}
& EMADS (3D U-Net)
& 0.793 & 0.657 & 0.563 & 0.780 & 2.653 & 0.841 & 1.812 \\

& EMADS (Swin UNETR)
& 0.810 & 0.681 & 0.692 & 0.729 & 2.329 & 0.708 & 1.621 \\

& SomaNet (DINOv2) [Ours]
& \textbf{0.991} & \textbf{0.982} & 0.774 & 0.078 & 0.395 & 0.203 & 0.192 \\

& SomaNet (SwinIR) [Ours]
& 0.990 & 0.980 & \textbf{0.870} & \textbf{0.044} & \textbf{0.285} & \textbf{0.148} & \textbf{0.137} \\

\midrule

\multirow{4}{*}{\textbf{Average}}
& EMADS (3D U-Net)
& 0.814 & 0.687 & 0.603 & 0.758 & 2.535 & 0.847 & 1.688 \\

& EMADS (Swin UNETR)
& 0.776 & 0.635 & 0.659 & 0.776 & 2.526 & 0.682 & 1.844 \\

& SomaNet (DINOv2) [Ours]
& \textbf{0.992} & \textbf{0.984} & 0.740 & 0.091 & 0.403 & 0.224 & 0.179 \\

& SomaNet (SwinIR) [Ours]
& 0.990 & 0.981 & \textbf{0.818} & \textbf{0.074} & \textbf{0.369} & \textbf{0.184} & \textbf{0.185} \\

\bottomrule
\end{tabular}
}
\end{table*}

\subsection{Results}
SomaNet is evaluated quantitatively and qualitatively on the two held-out EMADS test blocks. We further demonstrate its ability to propagate sparse annotations by generating dense 3D instance segmentations on partially annotated training volumes.

\subsubsection{Results on Test Data}
\begin{figure*}[!ht]    \centering
    \includegraphics[width=0.999\linewidth]{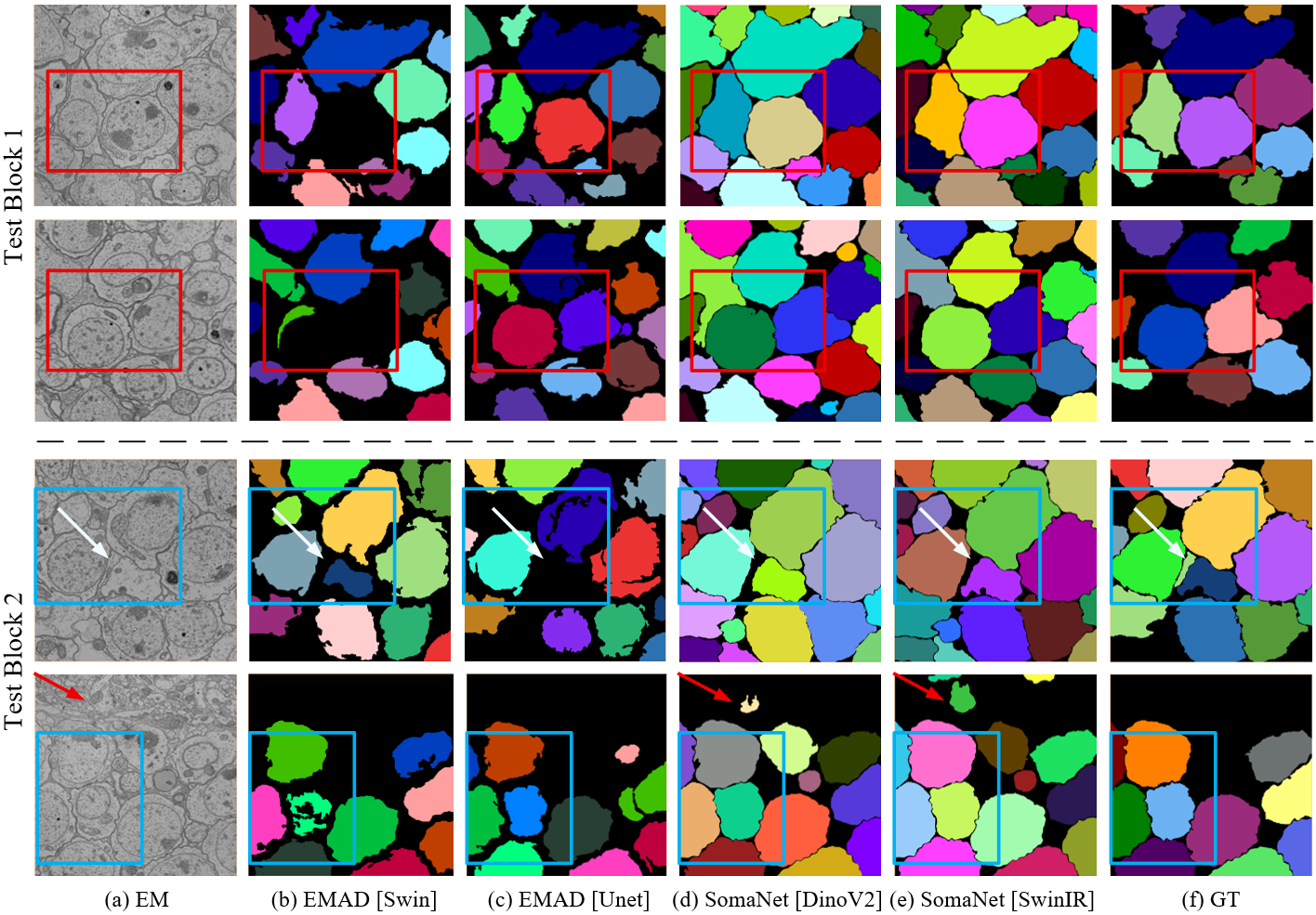}
    \caption{Qualitative comparison of the proposed SomaNet and the EMADS baselines on the two test sets. Columns show (a) input EM image, (b) EMADS (Swin UNETR), (c) EMADS (3D U-Net), (d) SomaNet (DINOv2), (e) SomaNet (SwinIR), and (f) ground-truth annotation. Across both test sets, SomaNet produces more accurate, boundary-preserving, and topologically consistent soma instance segmentations, particularly in densely packed regions. Nevertheless, it occasionally misses very small soma cells (white arrow) in Test Block 2.}
    \label{fig:testset_1_and_2}
\end{figure*}

SomaNet is evaluated on the two held-out EMADS test blocks. Quantitative comparisons are summarized in Table~\ref{tab:method_comparison}, while representative qualitative results are shown in Fig.~\ref{fig:testset_1_and_2}.
Across both test blocks, SomaNet consistently outperforms the EMADS baselines across all evaluation metrics. The proposed framework achieves substantially higher Dice, IoU, and PQ while simultaneously reducing ARAND and VOI, indicating improved semantic overlap, more accurate instance delineation, and fewer merge and split errors. These improvements are consistently observed across both backbone architectures, demonstrating that the performance gains primarily arise from the proposed affinity-based learning framework rather than from a particular network architecture.

Among the evaluated variants, the SwinIR backbone achieves the best overall instance segmentation performance, obtaining an average PQ of $0.818$, ARAND of $0.074$, and VOI of $0.369$, while maintaining an average Dice of $0.990$ and IoU of $0.981$. The DINOv2 variant achieves comparable performance despite using a largely frozen pretrained foundation-model encoder, attaining the highest average Dice ($0.992$) and IoU ($0.984$), together with a PQ of $0.740$, ARAND of $0.091$, and VOI of $0.403$. These results suggest that the dense image representations learned by SwinIR are particularly well suited for affinity-based instance delineation, whereas the semantically rich features of DINOv2 provide slightly stronger foreground localization. Nevertheless, the competitive performance of both backbones demonstrates the architectural flexibility of SomaNet and its ability to effectively leverage both task-specific networks and modern foundation models.

Compared with the EMADS Swin UNETR baseline, SomaNet with SwinIR improves the average PQ from $0.659$ to $0.818$ and decreases the average VOI from $2.526$ to $0.369$. It also reduces the average ARAND to $0.074$, compared with $0.776$ for Swin UNETR and $0.758$ for the 3D U-Net baseline, demonstrating substantially more accurate and consistent 3D soma instance segmentation.

The qualitative results further corroborate these findings. As illustrated in Fig.~\ref{fig:testset_1_and_2}, SomaNet produces smooth, anatomically consistent soma instance segmentations while maintaining clear separation between adjacent cells. In contrast, the EMADS baselines frequently produce fragmented segmentations with incomplete boundaries and artificial gaps between densely packed neighboring somata.

The first two rows correspond to representative examples from Test Block~1. In the first example (top row), the Swin UNETR baseline fails to recover the centrally located soma, whereas the 3D U-Net baseline introduces artificial gaps between densely packed neighboring somata. In contrast, both SomaNet variants successfully recover complete soma instances with accurate boundary separation (red boxes). In the second example, both EMADS baselines exhibit incomplete soma delineation. While both SomaNet variants substantially improve the reconstruction quality, the DINOv2 backbone recovers a slightly larger extent of the elongated soma than the SwinIR variant (blue boxes).

The last two rows show representative examples from Test Block~2. Similar to Test Block~1, both SomaNet variants consistently produce more accurate and topologically consistent segmentations than the EMADS baselines, particularly in the regions highlighted by the blue boxes. Both variants fail to recover a very small portion of one soma (third row, white arrow), indicating that extremely fine structures remain challenging for all evaluated methods. In the fourth row, both SomaNet variants accurately delineate the large soma; however, while the SwinIR variant successfully segments a small neighboring soma, the DINOv2 variant recovers it only partially (red arrow).

\subsubsection{Results on Partially Annotated Training Data}

\begin{figure*}[!ht]
    \centering
    \includegraphics[width=0.95\linewidth]{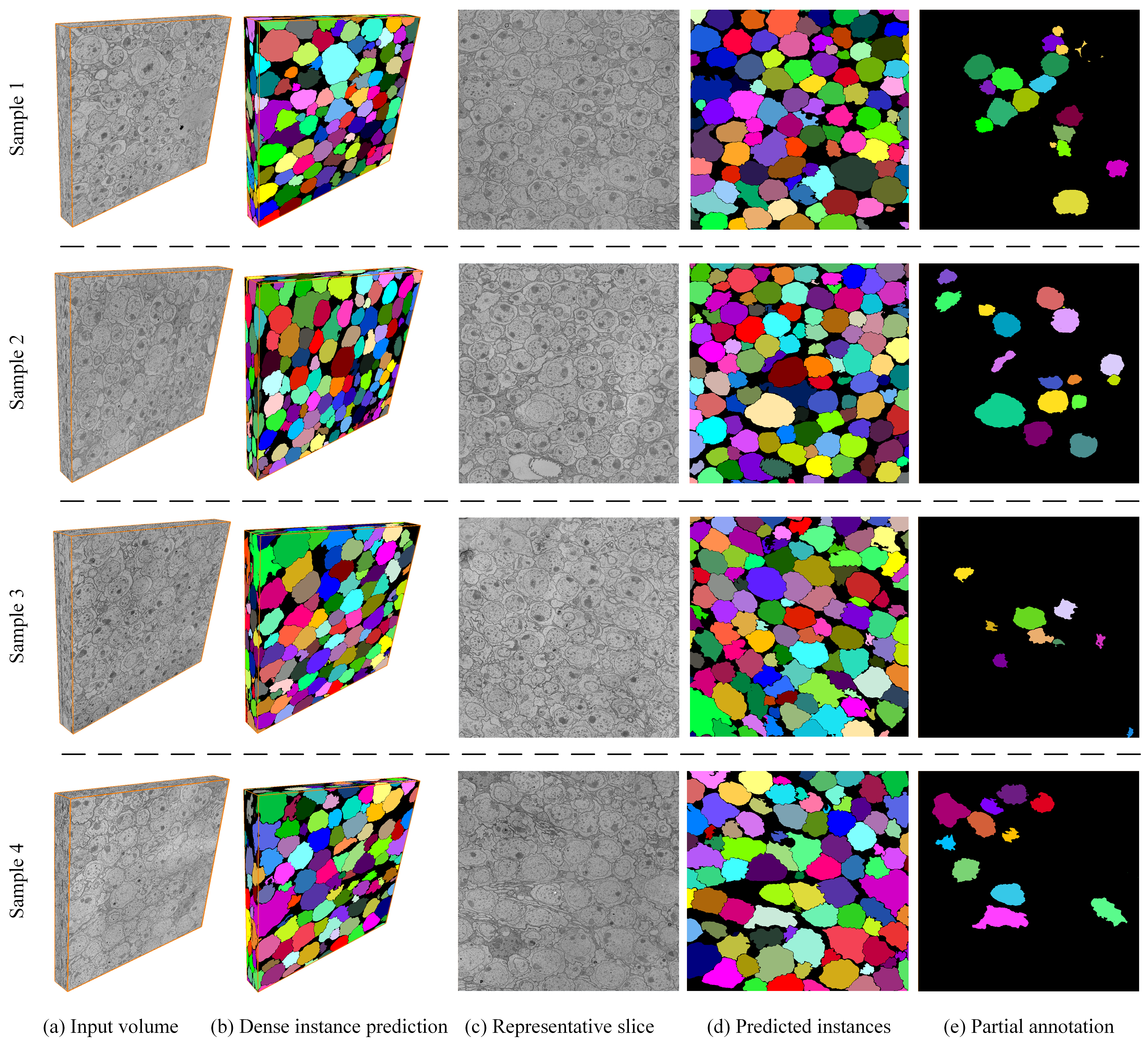}
    \caption{Qualitative results on four representative training volumes selected from the partially annotated EMADS training dataset. From left to right: (a) input 3D EM volume, (b) dense 3D soma instance segmentation generated by SomaNet, (c) representative EM slice, (d) predicted instance labels on the corresponding slice, and (e) available partial ground-truth annotations used for supervision. Despite supervision from only a subset of annotated soma instances, SomaNet successfully propagates the available annotations to recover dense, boundary-accurate, and topologically consistent 3D soma instance segmentations across volumes exhibiting diverse soma distributions, morphologies, and imaging characteristics.}
    \label{fig:partial_to_dense}
\end{figure*}

Figure~\ref{fig:partial_to_dense} presents qualitative results on four representative partially annotated training volumes exhibiting diverse soma densities, morphologies, and imaging characteristics. Despite being trained with annotations for only a subset of soma instances, SomaNet successfully propagates the available supervision to recover dense and accurate 3D soma instance segmentations, reconstructing previously unlabeled soma instances while preserving accurate boundaries and instance identities. Notably, many correctly segmented somata lie outside the annotated regions, which demonstrates that the SomaNet framework effectively generalizes beyond the available supervision rather than simply reproducing the labeled instances. The consistent performance across heterogeneous training volumes indicates that SomaNet can substantially reduce the manual annotation effort required for large-scale 3D EM connectomic reconstruction.

\subsubsection{Robustness to Imaging Artifacts}
\begin{figure}[!t]
    \centering
    \includegraphics[width=0.93\linewidth]{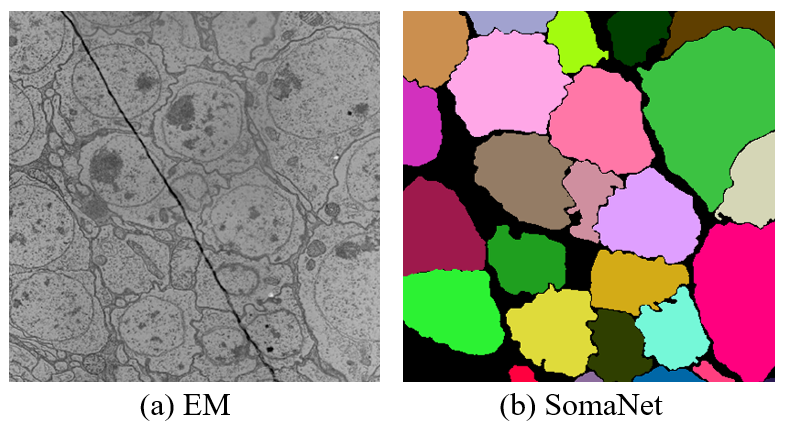}
    \caption{Robustness of SomaNet under challenging ssTEM imaging conditions. (a) Representative EM image with a characteristic line artifact and low-contrast soma boundaries. (b) Corresponding SomaNet prediction, accurately delineating individual soma instances while preserving clear separation between adjacent cells.}
    \label{fig:robustness}
\end{figure}

Figure~\ref{fig:robustness} demonstrates the robustness of SomaNet under challenging ssTEM imaging conditions. Despite characteristic line artifacts, which interrupt image continuity and obscure soma boundaries, together with low-contrast cell boundaries, the proposed framework accurately delineates individual soma instances while preserving clear separation between adjacent cells.

\subsubsection{Computational Time}

The average inference time of SomaNet is approximately 22 minutes per test volume, comprising about 20 minutes for slice-wise inference over 550 EM slices (each of size $650 \times 650$ pixels) and approximately 2 minutes for 2D-to-3D instance reconstruction. In comparison, the 3D U-Net and Swin UNETR baselines require approximately 2 minutes per volume. Although SomaNet incurs higher computational cost, it consistently achieves substantially superior instance segmentation performance, representing a favorable trade-off between computational efficiency and reconstruction quality.

\begin{figure}[!t]
    \centering
    \includegraphics[width=0.93\linewidth]{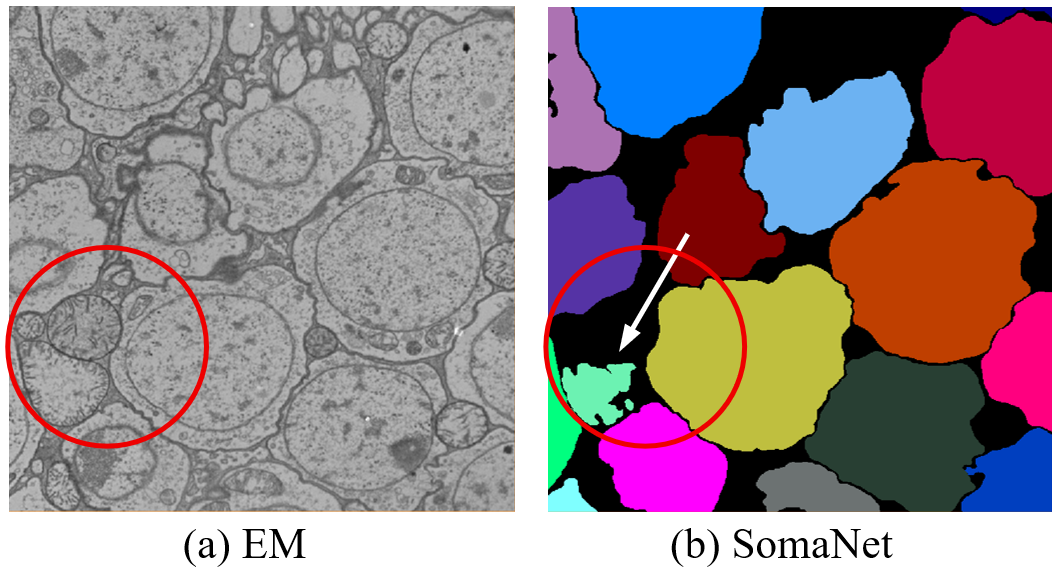}
    \caption{Representative failure case of SomaNet. The white arrow indicates a false-positive soma detection.}
    \label{fig:failure}
\end{figure}

\subsubsection{Limitations and Failure Mode Analysis}
Despite its overall effectiveness, SomaNet exhibits several failure modes. Non-soma structures with appearances and boundaries similar to cell somas may occasionally result in false-positive detections (white arrow in Fig.~\ref{fig:failure}). In addition, cell somas with elongated or highly irregular morphologies may be only partially recovered, resulting in incomplete segmentations (red box, second row, in Fig.~\ref{fig:testset_1_and_2}). In rare cases, very small cell soma instances may be missed or only partially recovered, particularly when their boundaries are indistinguishable from the surrounding tissue, as illustrated by the white arrow in the third row of Fig.~\ref{fig:testset_1_and_2}. These limitations are not unique to SomaNet and are also observed in competing methods, which shows the inherent difficulty of these challenging cases. Future work could incorporate explicit hard-negative mining and contrastive learning to further improve discrimination between cell somas and visually similar non-soma structures.


\section{Conclusion}
We presented SomaNet, a weakly supervised framework for soma instance segmentation in 3D EM under partial annotation. The proposed framework combines teacher--student learning with affinity-based supervision to recover dense and accurate instance segmentations from incomplete annotations. By integrating semantic-guided affinity decoding with geometric 2D-to-3D reconstruction, SomaNet preserves the large receptive field and strong in-plane representations of 2D backbones while producing volumetrically consistent 3D soma instances. Our theoretical analysis further shows that conventional mask supervision introduces systematic bias under partial annotation, whereas affinity learning is less susceptible to this bias, resulting in more reliable boundary supervision. The architecture-flexible design enables seamless integration of diverse backbone networks, including modern foundation models, facilitating the transfer of pretrained visual representations to 3D EM segmentation. Experiments on 3D EM brain volumes demonstrate that accurate and topology-preserving soma instance segmentation can be achieved with substantially reduced annotation effort. Overall, SomaNet provides a practical and scalable framework for annotation-efficient connectomic reconstruction without requiring exhaustive manual annotation.
\appendix
\subsection{Learning under Partial Annotations}
\label{app:affinity_robustness}

Under partial annotation, only a subset of voxels $\Omega \subset \mathcal{V}$ 
is labeled, while the remaining voxels $\Omega^u = \mathcal{V} \setminus \Omega$ 
are unlabeled and treated as background targets in the loss during supervised training. Let $y(v) \in \{0,1\}$ denote the \emph{latent} true foreground label at voxel $v$, and let $\hat{y}_\theta(v)\in(0,1)$ denote the model prediction.

\textbf{Mask loss bias under partial annotation.}
The fully supervised mask loss, decomposed over labeled and unlabeled regions, is
\begin{equation}
\mathcal{L}_{\mathrm{mask}}^{\mathrm{full}}(\theta) = \frac{1}{|\mathcal{V}|} 
\left(\sum_{v\in\Omega}\mathrm{CE}(y(v),\hat{y}_\theta(v)) + 
\sum_{v\in\Omega^u}\mathrm{CE}(y(v),\hat{y}_\theta(v))\right).
\end{equation}
The partially supervised mask loss, where unlabeled voxels are treated as background, is
\begin{equation}
\label{eq:partial_label}
\mathcal{L}_{\mathrm{mask}}^{\mathrm{partial}}(\theta) = \frac{1}{|\mathcal{V}|}
\left(\sum_{v\in\Omega}\mathrm{CE}\!\left(y(v),\hat{y}_\theta(v)\right) +
\sum_{v\in\Omega^u}\mathrm{CE}\!\left(0,\hat{y}_\theta(v)\right)\right).
\end{equation}
Subtracting the two objectives and taking gradients yields
\begin{equation}
\label{eq:mask_grad_decomp}
\begin{split}
\nabla_\theta \mathcal{L}_{\mathrm{mask}}^{\mathrm{partial}} &-  
\nabla_\theta \mathcal{L}_{\mathrm{mask}}^{\mathrm{full}} \\
&=  \frac{1}{|\mathcal{V}|} \sum_{v \in \Omega^u}
\nabla_\theta \Big(\mathrm{CE}(0,\hat{y}_\theta(v)) - 
\mathrm{CE}(y(v),\hat{y}_\theta(v))\Big). 
\end{split}
\end{equation}

For any unlabeled foreground voxel $v \in \Omega^u$ with $y(v)=1$, the 
per-voxel residual reduces to
\begin{equation}
\mathrm{CE}(0,\hat{y}_\theta(v)) - \mathrm{CE}(1,\hat{y}_\theta(v)) 
= \log\frac{\hat{y}_\theta(v)}{1-\hat{y}_\theta(v)}.
\end{equation}

Let $z_\theta(v)$ denote the corresponding logit such that 
$\hat{y}_\theta(v)=\sigma(z_\theta(v))$. Using
$\frac{\partial\,\mathrm{CE}(t,\sigma(z))}{\partial z} = \sigma(z) - t$,
we obtain
\begin{equation}
    \frac{\partial}{\partial z_\theta(v)}
    \Big(
    \mathrm{CE}(0,\hat{y}_\theta(v)) - \mathrm{CE}(1,\hat{y}_\theta(v))
    \Big) = 1 > 0.
\end{equation}

Thus, for unlabeled foreground voxels, the residual term contributes a strictly 
positive derivative with respect to the logit. Under gradient descent, this 
decreases $z_\theta(v)$ and therefore $\hat{y}_\theta(v)$, biasing predictions 
toward background. Consequently, whenever 
$\Pr(v \in \Omega^u,\; y(v)=1) > 0$, the gradient induced by 
$\mathcal{L}_{\mathrm{mask}}^{\mathrm{partial}}$ is generally biased relative to the fully supervised gradient, leading to false negatives at unlabeled foreground locations.

\textbf{Affinity loss under partial annotation.}
Let $\Omega \subset \mathcal{V}$ denote annotated voxels and $\Omega^u = \mathcal{V}\setminus\Omega$ the unannotated set. Unannotated voxels are assigned label $0$, and affinities are defined over all spatially valid pairs $(v,k)$ as $A_{v,k}=\mathbbm{1}[y(v)=y(k)]$. Let $A^*_{v,k}=\mathbbm{1}[y^*(v)=y^*(k)]$ denote the latent true affinity. Errors arise only when at least one of $v,k$ is an unannotated foreground voxel, i.e., when $A_{v,k} \neq A^*_{v,k}$. The affinity loss in Eq.~(\ref{eq:aff_loss}) yields
\begin{equation}
\frac{\partial \mathcal{L}_{\mathrm{aff}}}{\partial \hat{A}_{v,k}} \propto (\hat{A}_{v,k}-A_{v,k}),
\end{equation}
so the gradient sign determines whether embeddings are pulled together or pushed apart.

Unlike the mask loss (Eq.~(\ref{eq:partial_label})), which induces a uniformly signed gradient on unlabeled foreground voxels, the affinity-induced gradients have mixed signs: some erroneous pairs push embeddings apart, while others pull them together. As a result, the errors remain local and partially cancel rather than accumulating systematically. Hence, affinity supervision is less biased to partial annotation, although not entirely bias-free.

\vspace{-7mm}
\section*{Acknowledgment}
The authors thank CSC–IT Center for Science, Finland, and the Bioinformatics Center of the University of Eastern Finland for computational resources.
\vspace{-3mm}
\bibliographystyle{ieeetr}
\bibliography{main.bib}

\end{document}